\documentclass[twocolumn,showpacs,preprintnumbers,amsmath,amssymb,aps,prl,groupedaddress]{revtex4-1}

\usepackage{graphicx} % Include figure files
\usepackage{subfigure}
\usepackage{dcolumn} % Align table columns on decimal point
\usepackage{bm} % bold math
\usepackage{color}
\usepackage{ulem}
\usepackage{xfrac}
\usepackage{blindtext}
\definecolor{linkcolor}{rgb}{0,0,0.6}		% d?finition de la couleur des liens pdf
\usepackage[	pdftex,colorlinks=true,	
			pdfstartview=FitV,
			linkcolor= linkcolor,
			citecolor= linkcolor,
			urlcolor= linkcolor,
			hyperindex=true,
			hyperfigures=false]
			{hyperref}				% fichiers pdf 'intelligents', avec des liens entre les r?f?rences, etc.
\usepackage[francais]{babel}
\usepackage[latin1]{inputenc}       	% utilisation directe des caract?res accentu?s sur mac
\usepackage{cancel}

\begin{document}

\title{Experimental evidence of thermal-like behaviour in dense granular suspensions}

\author{Nariaki Saka\"i\textsuperscript{1}}
\email{nariaki.sakai@lps.ens.fr}
\author{S\'ebastien Moulinet\textsuperscript{1}}
\author{Fr\'ed\'eric Lechenault\textsuperscript{1}}
\author{Mokhtar Adda-Bedia\textsuperscript{2}} 
\affiliation{\textsuperscript{1}Laboratoire de Physique Statistique, Ecole Normale Sup\'erieure, PSL Research University, Sorbonne University, CNRS, F-75231 Paris, France}
\affiliation{\textsuperscript{2}Universit\'e de Lyon, Ecole Normale Sup\'erieure de Lyon, Universit\'e Claude Bernard, CNRS, Laboratoire de Physique, F-69342 Lyon, France}

\date{\today}

\begin{abstract}
We experimentally investigate the statistical behaviour of a model two-dimensional granular system undergoing stationary sedimentation. Buoyant cylindrical particles are rotated in liquid-filled drum, thus confined in a harmonic centripetal potential with tunable curvature, which competes with gravity to produce various stationary states: though heterogeneous, the packing fraction of the system can be tuned to be fully dispersed to fully crystallised as the rotation rate is increased. We show that this dynamical system is in mechanical equilibrium in the confining potential and exhibits a thermal-like behaviour, where the granular pressure and the packing fraction are related through an equation of state.  We obtain a semi-analytical expression of the equation of state allowing to probe the nature of the hydrodynamic interactions between the particles. This description is valid in the whole range of the physical parameters we investigated and reveals a buoyant energy scale that we interpret as an effective temperature. We finally discuss the behaviour of our system at high packing fractions and the relevance of the equation of state to the liquid-solid phase transition.
\end{abstract}

\maketitle

Statistical approaches to the phase behaviour of granular matter have flourished during the past years, from kinetic theories to Edwards hypothesis, culminating with the jamming paradigm~\cite{liu2010jamming}. However, no unifying framework has yet emerged that captures the physics of this class of systems in the same the way as thermal statistics for molecular systems. Energy dissipation and athermality are two defining features of granular matter: since thermal fluctuations are irrelevant for millimeter-sized particles, achieving a dynamical steady state requires a continuous energy injection to compensate for the dissipative processes. This usually takes the form of a mechanical agitation which plays the role of the thermal bath. When this is achieved, particles behave like a fluid, and for 2D monodisperse systems, the granular fluid crystallises when the density of particles is increased~\cite{Strassburger2000, Wu2005, Reis2006, Reyes2008, Komatsu2015}. These observations are very similar to what is observed in simulations of hard disks with elastic collisions~\cite{Qi2014b,Bernard2011}, or for colloidal systems~\cite{Marcus1996, Keim2007, Wang2010, Dillmann2012}, despite the fact that granular fluids are out of equilibrium. However, the depth of this analogy remains elusive, partly due to dissipative processes like solid friction, \textit{e.g.} the so-called granular temperature does not equilibrate between phases when there is coexistence~\cite{Prevost2004}. This leads to question to what extent concepts from thermodynamics can be exported to these out of equilibrium situations~\cite{DAnna2003, Ojha2004, Castillo2012, Puckett2013, Luu2013a}.

On the one hand, most of the experimental studies on dense granular media have been carried out by controlling the packing fraction, \textit{e.g.} by changing the number of particles in a given fixed volume with hard walls~\cite{Reis2006}. On the other hand, simple granular sedimentation experiments cannot be sustained within dynamical steady-states for long enough to decipher the resulting statistical ensemble they evolve in. Here we present an experimental situation where a buoyant granular suspension is instead confined in a harmonic trap with tunable curvature and maintained in a continuous sedimentation state. In addition to gravity effects induced by the mismatch in density between the fluid and the grains, a centripetal confining pressure is adjusted by changing the rotation rate of the system. The competition between this confinement and buoyancy allows us to select the density profile of the assembly and thus to explore various packing states. Furthermore, instead of solid friction, the particles are coupled through hydrodynamic interactions. Altogether, the system can be continuously driven from a dispersed to a crystalline state with various spatial profiles of the packing fraction.

In this Letter, we study the statistical properties of such suspension as a function of the confinement for a large range of values of density contrasts between the grains and the solution. The spatial distribution of particles can be approached using a functional with a unique fitting parameter. Using the fact that the suspension is in a dynamical stationary state at mechanical equilibrium, we define a granular pressure and relate it implicitly to the local packing fraction through a single energy scale. The identification of the corresponding shape function enables to explicit the equation of state and to define an effective temperature of the suspension. This effective temperature is shown to scale with a gravitationnal energy scale and the square root of the density contrast, thus mixing buoyant and inertial effects. 

\begin{figure}[htb]
\centering
\includegraphics[width=\linewidth]{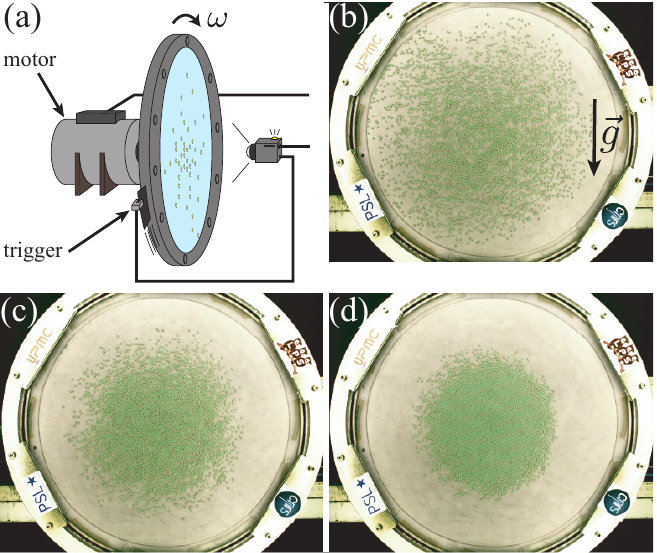}
\caption{(a) Schematics of the experimental setup. (b)-(d) Snapshot of the system for a relative density contrast of $\Delta\rho/\rho = 5.4\%$ at rotation frequencies $0.2\mathrm{Hz}$ (b), $0.4\mathrm{Hz}$ (c) and $0.6\mathrm{Hz}$ (d).}
\label{fig:fig1}
\end{figure}

The experimental setup displayed in Fig.~\ref{fig:fig1}a is inspired from~\cite{Bayart2014}. It consists in a monolayer of $N$ cylindrical particles of diameter $d=4\mathrm{mm}$, height $h = 2\mathrm{mm}$, mass $m=25\mathrm{mg}$ and density $\rho=1.053 \mathrm{g.cm^{-3}}$ in a two-dimensional cylindrical drum of radius $250\mathrm{mm}$. The two parallel plates of the drum are separated by a distance $1.5h$ and the cell is filled with a solution of cesium chloride, making the liquid denser than the particles ($\rho_l>\rho$). We explore granular assemblies with $3000\le N\le 3800$ in solutions with relative density contrasts in the range $0.38\%<\Delta\rho/\rho<71\%$. The axis of the cell is horizontal, so particles undergo the effect of gravity, and the system is rotated with a motor at a frequency $f$ ranging from $0.07$Hz to $1.70$Hz. A high resolution camera is placed in front of the cell and triggered by means of an optical fork once every cycle. We focus on the statistics of the assembly computed over a series of $2000$ pictures per value of the rotation rate. Particles are hollow cylinders made of white polystyrene and filled with a green silicon core to ease their detection; the error on the positions is of the order of $30\mu m$.

The phenomenology of this system is quite rich: at very low rotation rate, the grains float up to the top of the cell and avalanche similarly to what occurs in a partially filled rotating drum~\cite{Sepulveda2005}, a regime we do not study here. Three pictures of the experiment at higher rotation frequencies are shown in Figs.~\ref{fig:fig1}(b-d); they correspond respectively to the fully dispersed state where the inter-particle distance is larger than $d$, the state near the critical point where a dense, disordered region has pervaded a significant central region, and finally a state with a large ordered crystal in the center of the cell surrounded by a gas-like ring. An illustrative movie of the different phases is provided in~\footnote{See Supplemental Material at [URL will be inserted by publisher] for movies illustrating the phase behaviour of the experiment}. It is noteworthy that the system is quite isotropic in the azimuthal direction with respect to the center of mass of the assembly, a property that we will use in the subsequent analysis. However, gravity appears to have a mixing effect on the assembly.

\begin{figure}[htb]
\includegraphics[width=\linewidth]{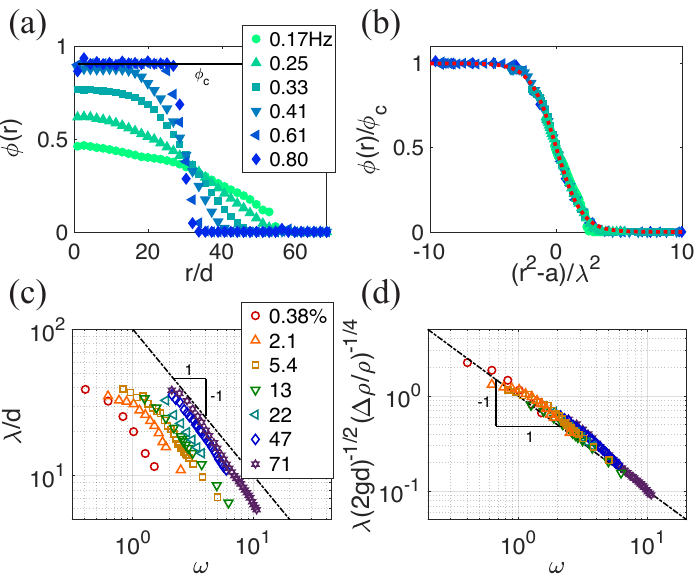}
\caption{(a) Radial packing fraction for selected rotation frequencies $0.17\mathrm{Hz}<f<0.8\mathrm{Hz}$ at $\Delta \rho/\rho = 5.4\%$. The horizontal line locates the maximum fully crystalline packing fraction $\phi_c$. (b). Corresponding rescaled packing fraction profiles using the fitting function given by Eq.~(\ref{eq:densityFit}). (c) The fitting parameter $\lambda$ as a function of $\omega=2\pi f$ and for different density contrasts. (d) Plot of $\lambda$ rescaled using Eq.~(\ref{eq:lambda}).}
\label{fig:fig2}
\end{figure}

To characterise the spatial distribution of particles in the cell, we first compute the mean packing fraction $\phi(r)=\left\langle dA(r)/2\pi rdr\right\rangle$ where $dA(r)$ is the cumulated area of the particles located in a ring of width $dr$ at distance $r$ from the instantaneous centre of mass of the grains, which does not necessarily coincide with the center of the cell. Fig.~\ref{fig:fig2}(a) shows that at low frequency, $\phi(r)$ is roughly parabolic. However, at larger frequencies, the packing fraction starts developing a plateau at the center, corresponding to a dense region where particles are in contact. At further higher rotation rates, the value of the plateau tends to the packing fraction of the hexagonal crystal $\phi_c=\frac{\pi}{\sqrt{12}}$ as the assembly fully orders.

To gain insight on the generic behaviour of the packing fraction with experimental control parameters $f$, $\Delta \rho/\rho$, $d$ and $N$, we search for a possible functional form of $\phi(r)$. First, to respect the rotational symmetry and cope with the saturation at small distance and the absence of particles at large distance, the profiles of density distributions are adjusted using the following shape:
\begin{equation}
\phi(r) = \frac{\phi_c}{e^{\frac{r^2-a}{\lambda^2}} + 1}\;,
\label{eq:densityFit}
\end{equation}
where $a$ and $\lambda$ are fitting constants. Also, taking advantage that for each experimental run the number of particles $\pi d^2 N/4=\int_0^\infty \phi(r) 2\pi rdr$ is fixed, the parameter $a$ can be fixed through:
\begin{equation}
a = \lambda^2 \log\left(e^{\frac{R^2}{\lambda^2}}-1\right)\;;\qquad R= \frac{d}{2}\sqrt{\frac{N}{\phi_c}}\;,
\label{eq:a}
\end{equation}
where $R$ is the radius of the fully crystallised granular assembly. Notice that Eq.~(\ref{eq:a}) shows that $a$ can be either positive or negative depending on whether $R/\lambda$ is larger or smaller than $\sqrt{\log 2}$. The rescaled profiles depicted in Fig.~\ref{fig:fig2}(b) confirm the relevance of density profiles given by Eq.~(\ref{eq:densityFit}). We are thus left with a single fitting parameter $\lambda$ that is determined as function of the experimental parameters. Fig.~\ref{fig:fig2}(c) shows that this characteristic length scale is inversely proportional to the rotation rate $\omega=2\pi f$ for all density contrasts. Moreover, we notice that normalising $\lambda$ by the square root of the density contrast collapses all dependance on a master curve exhibiting a $\omega^{-1}$ dependence. Finally, Fig.~\ref{fig:fig2}(d) shows that $\lambda$ can be expressed in terms of the physical parameters of the system as
\begin{equation}
\lambda^2 =  \frac{2 g d}{\omega^2}\sqrt{\frac{\Delta \rho}{\rho}}\;.
\label{eq:lambda}
\end{equation}
where g is the standard gravity. Provided we measure density distributions in steady state conditions, and using the sole assumption of mechanical equilibrium, we can convert the measured profiles into a pressure measurement. In the sequel, we build a mean field model of a two-phase fluid in which the partial pressure of the particles verifies the simple hydrostatic equation:
\begin{eqnarray}
-\frac{\partial p}{\partial r} - \Delta \rho \omega^2 hr\phi(r) = 0\;,
\end{eqnarray}
which corresponds to the mechanical equilibrium of an inhomogeneous fluid under an external field \cite{barrat1992barometric, Ginot2015}. We can now integrate this expression - with $p(r\to\infty) = 0$ set by the absence of grains at $r\to\infty$ - and obtain the granular pressure as a function to the distance to the center of mass of the assembly:
\begin{equation}
p(r) = \Delta \rho \omega^2 h\int_{r(\phi)}^\infty r'\phi(r') \,\mathrm{d}r'\;.
\label{eq:Pmech}
\end{equation}
Since $\phi(r)$ is monotonic in $r$, we can invert this expression to obtain the pressure as a function of the local average packing fraction $\phi$. This yields the following equation of state:
\begin{equation}
\frac{\pi}{4} d^2\beta p(\phi)=-\phi_c\log \left(1-\frac{\phi}{\phi_c}\right)\;,
\label{eq:EOS}
\end{equation}
with
\begin{equation}
\beta^{-1}=\frac{1}{2}\Delta m\,\omega^2\lambda^2= \Delta m g d \left(\frac{\Delta \rho}{\rho}\right)^{\frac{1}{2}}\;,
\label{eq:beta}
\end{equation}
and $\Delta m=(\pi d^2h/4)\Delta \rho$ is the mass contrast of the displaced liquid volume and the grain.

\begin{figure}[htb]
\includegraphics[width=0.9\linewidth]{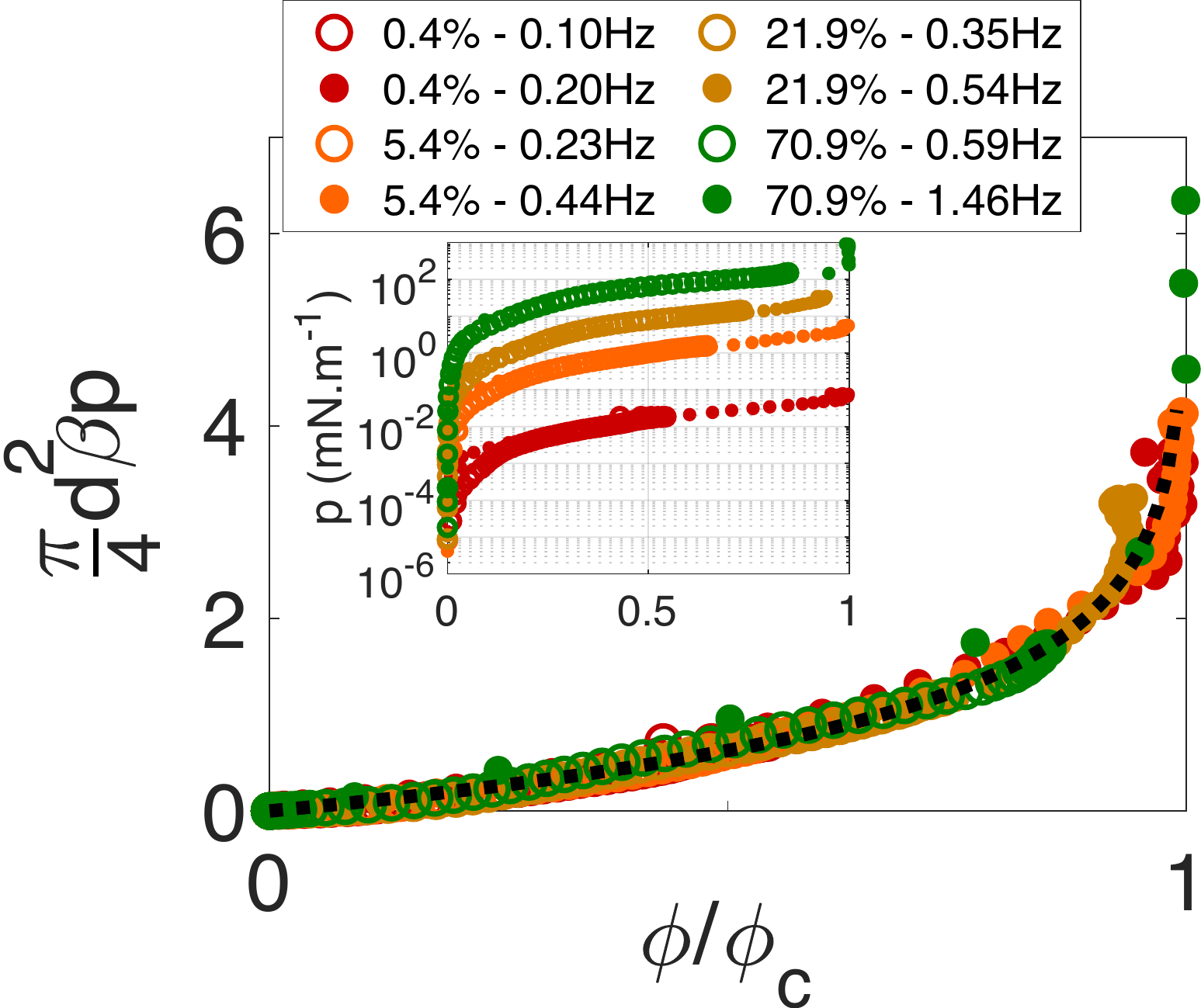}
\caption{Normalized granular pressure obtained from the mixture model for a range of frequencies and density contrasts. Black dotted line corresponds to the analytical expression for the pressure as given by Eq.~(\ref{eq:EOS}). The inset shows the non-normalized pressure in semi-log scale.}
\label{fig:fig3}
\end{figure}

The mechanical pressure as a function of the packing fraction is shown in Fig.~\ref{fig:fig3} for representative values of density contrasts and rotation rates. As all the curves surprisingly collapse without further rescaling, the normalised pressure appears to be independent of the rotation rate and thus of the confining potential. Interestingly, the prefactor $\beta$ can be interpreted as an effective temperature. Indeed, multiplying by $\frac{1}{2}\Delta m \omega^2$  both the numerator and the denominator in the exponential of the fitting function in Eq.~(\ref{eq:densityFit}) yields ``Fermi-Dirac-like" statistics:
\begin{equation}
\phi(r) = \frac{\phi_c}{e^{\beta\left(E(r)-\mu\right)}+1}\;,
\label{eq:FD}
\end{equation}
where $E(r) = \frac{1}{2} \Delta m \omega^2 r^2$ is the centripetal energy of a grain at distance $r$ from the center of mass of the assembly, $\beta$ is given by Eq.~(\ref{eq:beta}) and $\mu\equiv\frac{1}{2}\Delta m \omega^2 a$ appears as a chemical potential that insures conservation of the number of particles. This chemical potential can be simply related to the ``Fermi energy" $E_0\equiv \frac{1}{2}\Delta m \omega^2 R$ through $\exp\left(\beta\mu\right)=\exp\left(\beta E_0\right)-1$ (see Eq.~\ref{eq:a}). Notice that $\mu$ can be either positive or negative, meaning that bringing new grains into the system can either lower or increase its total energy. Eq.~(\ref{eq:beta}) shows that the effective temperature is independent of the rotation rate. $\beta$ depends only on a buoyant energy scale which means that the fluctuations are induced by the gravitational potential which injects energy into the rotating system. However the energy transfer is not perfect since it gives rise to a corrective prefactor $\sqrt{\frac{\Delta\rho}{\rho}}$ which originates in the competition of gravitational energy injection with inertia.

The expansion of Eq.~(\ref{eq:EOS}) for $\phi\rightarrow 0$ yields $\frac{\pi}{4} d^2\beta p(\phi)=\phi + \phi^2/(2\phi_c)+O(\phi^3)$, so that in the dilute limit the system behaves as an ideal gas. However, the second Virial coefficient is given by $B_2=\sqrt{3}/\pi<2$ suggesting that the system can not be described as purely repulsive hard disk but as a long range interacting particle liquid~\cite{Ginot2015}. The equation of state predicts a phase transition at density $\phi_c$ that corresponds to the fully compact crystalline phase. Close to this density the divergence of pressure is logarithmic in contrast to the behaviour in $(\phi_c-\phi)^{-1}$ predicted for hard spheres~\cite{torquato2010jammed}.

The rationale that leads to the surprisingly simple description of the states of our suspension introduced above relies on several salient experimental observations. The first one is that stationary states of our suspension are well defined: whatever the initial condition, the system tends to the same phase that solely depends on a fixed set of experimental parameters. This suggests that the dynamic is sufficiently ergodic to allow the system to explore its whole phase space, such that defining averaged quantities like mean packing fraction becomes of particular relevance. Moreover, the packing fraction profiles are rotationally symmetric, meaning that the competition between buoyency and centripetation is somehow dynamically balanced, which could be due to the strength of the hydrodynamic drag compared to the buoyancy.

Altogether, in this well controlled situation with no degenerate Coulomb contacts, a smooth confining potential and laminar flow, a thermodynamic-like description emmerges. The spatial distribution of particles comes from the competition between the confining energy and the buoyancy, which is expressed through the Boltzmann factor and the corresponding effective temperature. This description allows us to explicit an equation of state for the grains in suspension, that predicts a pressure divergence at the close-packed density $\phi_c$. Surprisingly, our system is insensitive to jamming, probably because we are looking at well-mixing stationary states without prior quenching procedure. 

Further investigation is needed to establish the relevance of this equation of state. First, the very existence of such thermodynamic description questions that of an underlying statistical mechanics, in particular in the presence of long-range interactions. We envision to test that relationship through careful measurements of the density fluctuations, which should relate to the compressibility through the usual fluctuation-dissipation relation. Second, the surprising result that the effective temperature only depends on the density contrast between the grains and the liquid might be seriously challenged in a situation where two kinds of grains with different densities are mixed; it is unclear what equilibration would mean in this context and is definitely worth testing. Finally, this experimental system is a good playground for probing many statistical aspects and the emergence of thermal-like properties in the stationary states of dynamical systems.

\textit{Acknowledgments} -- This work was carried out in the framework of the METAMAT project  ANR-14-CE07-0031 funded by Agence Nationale pour la Recherche.

\bibliographystyle{apsrev4-1}
\bibliography{PRL_Sakai_2018.bib}

\end{document}